\documentclass[preprint,amsmath,amssymb]{revtex4}
\usepackage{amssymb}
\usepackage{amsmath}
\usepackage{graphicx}
\usepackage{epsfig}
\usepackage{subfigure}
\usepackage{amsfonts}

\begin{document}
	
\title{Efficient scheme for realizing a multiplex-controlled phase gate with
	photonic qubits in circuit quantum electrodynamics}
\author{Qi-Ping Su$^{1}$}
\author{Yu Zhang$^{2}$}
\author{Liang Bin$^{1}$}
\author{Chui-Ping Yang$^{1,3}$}
\email{yangcp@hznu.edu.cn}

\address{$^1$School of Physics, Hangzhou Normal University, Hangzhou, Zhejiang 311121, China}
\address{$^2$School of Physics, Nanjing University, Nanjing 210093, China}
\address{$^3$Quantum Information Research Center, Shangrao Normal University, Shangrao, Jiangxi 334001, China}	
	
\begin{abstract}
We propose an efficient scheme to implement a multiplex-controlled phase
gate with multiple photonic qubits simultaneously controlling one target
photonic qubit based on circuit quantum electrodynamics (QED). For convenience, we denote this
multiqubit gate as MCP gate. The gate is realized by using a two-level
coupler to couple multiple cavities. The coupler here is a superconducting
qubit. This scheme is simple because the gate implementation requires only \textit{one step} of operation. In addition, this scheme is quite general because the two logic states of each photonic qubit can be encoded with a
vacuum state and an arbitrary non-vacuum state $\left\vert \varphi
\right\rangle $ (e.g., a Fock state, a superposition of Fock states, a cat
state, or a coherent state, etc.) which is orthogonal or quasi-orthogonal to
the vacuum state. The scheme has some additional advantages: Because only
two levels of the coupler are used, i.e., no auxiliary levels are utilized,
decoherence from higher energy levels of the coupler is avoided; the gate
operation time does not depend on the number of qubits; and the gate is
implemented deterministically because no measurement is applied. As an
example, we numerically analyze the circuit-QED based experimental
feasibility of implementing a three-qubit MCP gate with photonic qubits each
encoded via a vacuum state and a cat state. The scheme can be applied to
accomplish the same task in a wide range of physical system, which consists
of multiple microwave or optical cavities coupled to a two-level coupler
such as a natural or artificial atom.
\end{abstract}
\maketitle
\date{\today }

\section{Introduction and motivation}

Multiqubit gates (i.e., $n$-qubit gates with $n\geq 3$ ) are essential
elements in quantum networks, quantum simulation, and quantum information
processing (QIP). Generally speaking, there are two types of significant
multiqubit gates, which have drawn much attention during the past years. One
is a multiplex-controlled NOT or phase gate with multiple qubits
simultaneously controlling one target qubit (Fig. 1). The other is a multi-%
\textit{target}-qubit controlled NOT or phase gate with one qubit
simultaneously controlling multiple target qubits (Fig. 2). It is well known
that these two types of multiqubit gates are of significance in QIP. For
instance, they have applications in quantum algorithms [1-3], quantum
Fourier transform, error correction [4-6], quantum cloning [7], and
entanglement preparation [8].

A multiqubit gate can in principle be decomposed into a series of
single-qubit gates and two-qubit gates. However, with an increasing number
of qubits, the required number of single- and two-qubit quantum gates
increases drastically [9-12]. This means that based on the conventional gate
decomposition [9-12], the procedure for implementing a multiqubit gate is
complex and the gate is difficult to realize experimentally. Hence, it is
worth looking for efficient methods to \textit{directly} implement
multiqubit gates. 
\

\begin{figure}[tbp]
\begin{center}
\includegraphics[bb=14 116 900 434, width=11.0 cm, clip]{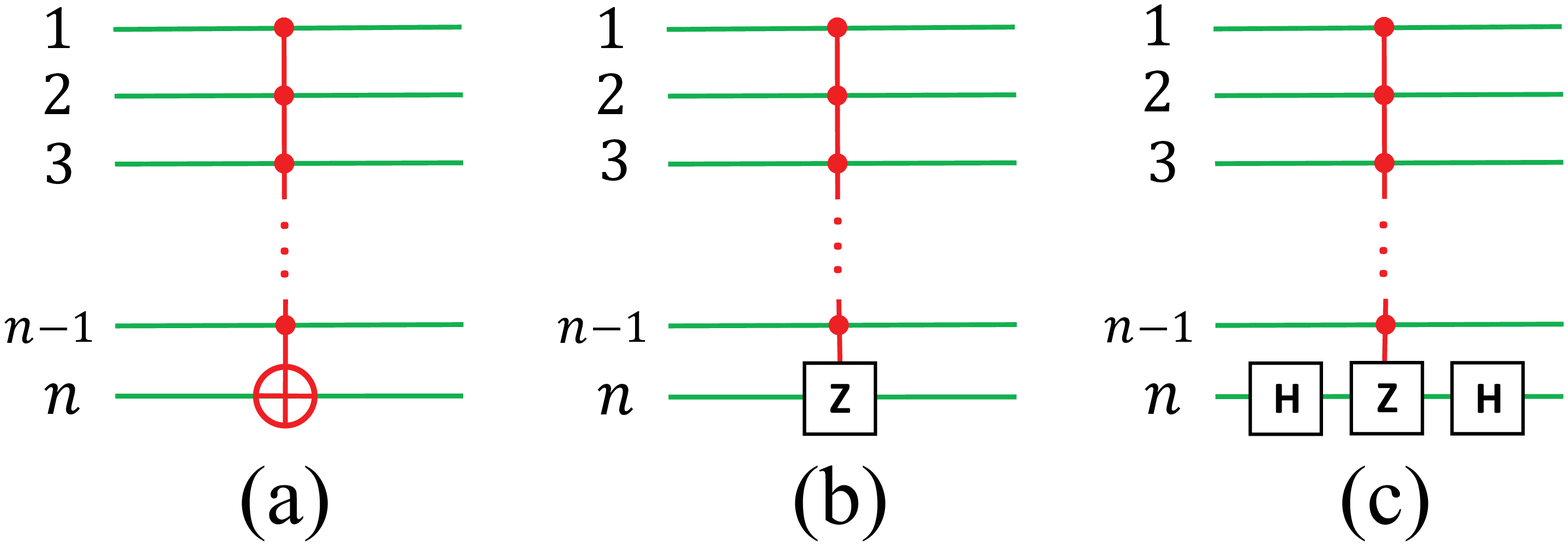} \vspace*{%
				-0.08in}
\end{center}
\caption{(color online) (a) Circuit of a multiplex-controlled NOT
gate (Toffoli gate) with $n-1$ control qubits ($1,2,...,n-1$) simultaneously
controlling a target qubit $n.$ When the $n-1$ qubits (on the filled circles)
are all in the state $\left\vert 1\right\rangle $, the state at $\oplus $
for the target qubit is bit flipped as $\left\vert 0\right\rangle
\rightarrow \left\vert 1\right\rangle $ and $\left\vert 1\right\rangle
\rightarrow \left\vert 0\right\rangle $. (b) Circuit of a
multiplex-controlled phase gate (MCP gate) with $n-1$ control qubits ($%
1,2,...,n-1$) simultaneously controlling a target qubit $n.$ When the $n-1$
qubits (on the filled circles) are all in the state $\left\vert
1\right\rangle $, the state $\left\vert 1\right\rangle $ at $Z$ for the
target qubit is phase flipped as $\left\vert 1\right\rangle \rightarrow
-\left\vert 1\right\rangle $ while nothing happens to the state $\left\vert
0\right\rangle $ at $Z$ for the target qubit. (c) Circuit for
constructing a Toffoli gate as shown in (a), by using a
MCP gate plus a single-qubit Hadamard gate on the
target qubit before and after the MCP gate. Here, H represents a Hadamard gate described by $\left\vert
0\right\rangle \rightarrow (1/\protect\sqrt{2})\left( \left\vert
0\right\rangle +\left\vert 1\right\rangle \right) $ while $\left\vert
1\right\rangle \rightarrow (1/\protect\sqrt{2})\left( \left\vert
0\right\rangle -\left\vert 1\right\rangle \right) $.}
\label{fig:1}
\end{figure}

\begin{figure}[tbp]
\begin{center}
\includegraphics[bb=64 115 850 478, width=9.5 cm, clip]{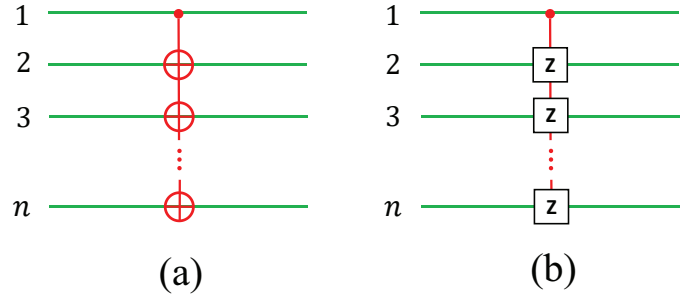} \vspace*{%
				-0.08in}
\end{center}
\caption{(color online) (a) Circuit of a multi-target-qubit
controlled NOT gate with one control qubit (qubit 1) simultaneously
controlling $n-1$ target qubits ($2,3,...,n$). If the control qubit (on the
filled circle) is in the state $|1\rangle$, then the state at $\oplus $ for
each target qubit is bit flipped as $\left\vert 0\right\rangle \rightarrow
\left\vert 1\right\rangle $ and $\left\vert 1\right\rangle \rightarrow
\left\vert 0\right\rangle $. (b) Circuit of a multi-target-qubit controlled phase gate with one
control qubit (qubit 1) simultaneously controlling $n-1$ target qubits ($%
2,3,...,n$). When the control qubit (on the filled circle) is in the state $%
\left\vert 1\right\rangle $, the state $\left\vert 1\right\rangle $ at $Z$
for each target qubit is phase flipped as $\left\vert 1\right\rangle
\rightarrow -\left\vert 1\right\rangle $ but nothing happens to the state $%
\left\vert 0\right\rangle$ at $Z$ for each target qubit.}
\label{fig:2}
\end{figure}

The focus of this work is on the implementation of the first type of
multiqubit gate. A multiplex-controlled NOT gate (Fig.1a), with multiple
qubits simultaneously controlling one target qubit, is often called as
Toffoli gate [13]. For convenience, a multiplex-controlled phase gate
(Fig.1b), with multiple qubits simultaneously controlling one target qubit,
is denoted as a\textit{\ MCP} gate throughout this paper. Over the past
years, a number of theoretical proposals have been put forward for \textit{%
directly} realizing a Toffoli gate or a MCP gate using \textit{matter}
qubits, such as trapped ionic qubits [14-16], quantum-dot qubits [17],
atomic qubits [18-25], NV-center qubits [26,27], and superconducting qubits
[28-31]. Experimentally, a three-qubit Toffoli gate or a three-qubit MCP
gate of matter qubits was demonstrated in NMR quantum systems [32],
superconducting qubits [33,34], or atomic systems [35]. Moreover, a
four-qubit Toffoli gate was experimentally implemented with superconducting
qubits [34]. On the other hand, theoretical proposals for realizing a
Toffoli gate or MCP gate with \textit{photons} have been presented by using
linear optical setups [36-40] and linear optical devices plus auxiliary
systems [41-43], and a three-qubit Toffoli gate of photons has been
experimentally demonstrated in a linear-optical setup [44,45].

Quite different from [14-45], schemes have also been proposed to \textit{%
directly} realize an $n$-qubit Toffoli or MCP gate of photonic qubits by a
cavity QED system [46] or a circuit QED system [47-49]. We note that the
previous works [46-49] only apply to the case that the two logic states of
each photonic qubit are encoded with a vacuum state and a single-photon
state. In addition, we note that the Toffoli or MCP gates discussed in Refs.
[46-49] were implemented essentially through \textit{step by step}
operations or by using a \textit{multi-level} natural or artificial atom to
couple multiple cavities. Generally speaking, decoherence from higher energy
levels could be a severe issue when a \textit{multi-level} quantum system is
employed in the gate realization; and \textit{step-by-step} operations are
not desirable in experiments, which increases the experimental complexity
and prolongs the operation time. For these reasons, in the following we will
propose a simple and more general scheme for the \textit{direct}
implementation of a MCP gate with photonic qubits based on circuit QED. The
circuit QED is analogue of cavity QED, it consists of superconducting (SC)
qubits and microwave resonators or cavities, and has been considered as one
of the most promising candidates for QIP [50-58]. In recent years, much
attention has been paid to the QIP with microwave photons, because microwave
photons can have lifetimes comparable to that of SC qubits [59].

As shown below, the present work works out for a more general case, i.e.,
the two logic states of each photonic qubit are encoded with a vacuum state
and an arbitrary non-vacuum state $\left\vert \varphi \right\rangle $ (e.g.,
a single-photon state, a Fock state, a superposition of Fock states, a cat
state, or a coherent state, etc.). The state $\left\vert \varphi
\right\rangle $ here is orthogonal or quasi-orthogonal to the vacuum state.
Moreover, the present work requires only a single-step operation and a
two-level coupler (a SC qubit)\ to couple $n$ cavities, thus decoherence
from higher energy levels is avoided in our gate realization and the gate
implementation procedure is greatly simplified when compared to [46-49]. To
the best of our knowledge, based on cavity QED or circuit QED, how to
realize a MCP gate of photonic qubits by using a \textit{two-level} coupler
and through only a \textit{single-step} operation has not been reported yet.

We stress that this work is on the implementation of a MCP gate with
multiple control qubits simultaneously controlling one target qubit (Fig.
1b), thus it is obviously different from the previous works (e.g.,
[20,60-68]) on the realization of a multi-\textit{target}-qubit gate with
one control qubit simultaneously controlling multiple target qubits (Fig.~2).

This paper is organized as follows. In Sec. II, we give a brief introduction
to the $n$-qubit MCP gate and the $n$-qubit Toffoli gate. In Sec. III, we
explicitly show how to realize an $n$-qubit MCP gate with $n$\ photonic
qubits each encoded via a vacuum state and a non-vacuum state. In Sec. IV,
we discuss the orthogonality required by the qubit encoding. In Sec. V, we
give a discussion on the circuit-QED based experimental feasibility of
implementing a three-qubit MCP gate with each photonic qubit encoded via a
vacuum state and a cat state, by using three one-dimensional (1D) microwave
cavities coupled to a SC flux qubit. A concluding summary is given in Sec.
VI.

\section{$n$-qubit MCP gate and Toffoli gate}

For $n$\ qubits, there exist $2^{n}$\ computational basis states, which form
a set of complete orthogonal bases in a $2^{n}$-dimensional Hilbert space of
the $n$\ qubits. An $n$-qubit computational basis state is denoted as $%
\left\vert i_{1}i_{2}...i_{n}\right\rangle $, where subscript $l$\
represents qubit $l$, and $i_{l}\in \{0,1\}$\ ($l=1,2,...,n$). The $n$-qubit
MCP gate considered in this work (Fig. 1b) is described as follows:

(i) When the $n-1$\ control qubits (say the first $n-1$\ qubits) are all in
the state $\left\vert 1\right\rangle $, the state $\left\vert 1\right\rangle
$\ of the target qubit (the last qubit) is phase flipped as $\left\vert
1\right\rangle \rightarrow -\left\vert 1\right\rangle $ (i.e., one has the
state transformation $\left\vert 11...1\right\rangle \rightarrow -\left\vert
11...1\right\rangle $ for the $n$ qubits), while the state $\left\vert
0\right\rangle $\ of the target qubit remains unchanged.

(ii) When even one of the $n-1$\ control qubits is not in the state $%
\left\vert 1\right\rangle ,$ nothing happens to both states $\left\vert
0\right\rangle $ and $\left\vert 1\right\rangle $ of the target qubit.

According to the description here, the $n$-qubit MCP gate can be
characterized by the following state transformation
\begin{eqnarray}
\left\vert 11...1\right\rangle &\rightarrow &-\left\vert 11...1\right\rangle
,  \notag \\
\left\vert i_{1}i_{2}...i_{n}\right\rangle &\rightarrow &\left\vert
i_{1}i_{2}...i_{n}\right\rangle ,\text{ for }\left\vert
i_{1}i_{2}...i_{n}\right\rangle \neq \left\vert 11...1\right\rangle ,
\end{eqnarray}
which shows that after the MCP gate, the computational basis state $%
\left\vert 11...1\right\rangle $ of the $n$ qubits changes to $-\left\vert
11...1\right\rangle ,$ while nothing happens to all other $2^{n}-1$\
computational basis states of the $n$ qubits.

The $n$-qubit Toffoli gate (Fig. 1a) is described by the following state
transformation:
\begin{eqnarray}
\left\vert 11...1\right\rangle _{12...n-1}\left\vert 0\right\rangle _{n}
&\rightarrow &\left\vert 11...1\right\rangle _{12...n-1}\left\vert
1\right\rangle _{n},  \notag \\
\left\vert 11...1\right\rangle _{12...n-1}\left\vert 1\right\rangle _{n}
&\rightarrow &\left\vert 11...1\right\rangle _{12...n-1}\left\vert
0\right\rangle _{n},
\end{eqnarray}
where the subscripts $1,2,...,n$ represent the $n$ qubits. Equation (2)
implies that when the $n-1$\ control qubits (the first $n-1$\ qubits) are
all in the state $\left\vert 1\right\rangle ,$ a bit flip (i.e., $\left\vert
0\right\rangle \rightarrow \left\vert 1\right\rangle $ and $\left\vert
1\right\rangle \rightarrow \left\vert 0\right\rangle $) happens to the
states $\left\vert 0\right\rangle $ and $\left\vert 1\right\rangle $ of the
target qubit (the last qubit). However, when even one of the $n-1$\ control
qubits is not in the state $\left\vert 1\right\rangle ,$ the states $%
\left\vert 0\right\rangle $ and $\left\vert 1\right\rangle $ of the target
qubit remain unchanged.

Since a Toffoli gate can be constructed with a MCP gate (Fig.~1c), in the
following we will mainly show how to implement the MCP gate (1) using
photonic qubits based on circuit QED.

\begin{figure}[tbp]
\begin{center}
\includegraphics[bb=295 247 670 409, width=11.5 cm, clip]{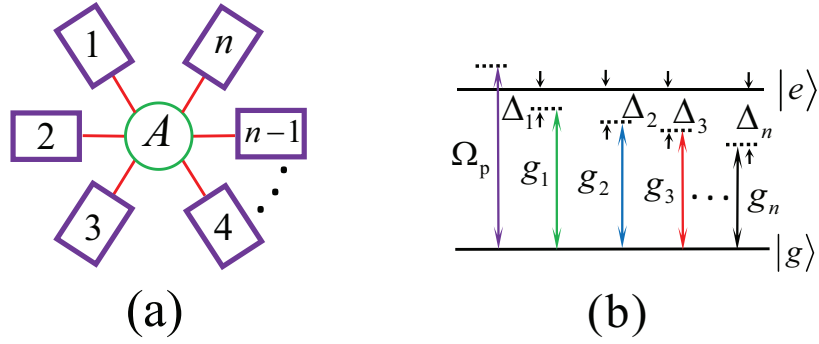} \vspace*{%
				-0.08in}
\end{center}
\caption{(color online) (a) Diagram of $n$ cavities coupled to a two-level
coupler (the circle $A$ in the middle). Here, the coupler is a two-level SC
qubit, which is capacitively or inductively coupled to each cavity. Each
square represents a cavity, which can be a one- or three-dimensional cavity.
(b) Illustration of $n$ cavities $(1,2,...,n)$ and a classical pulse
dispersively coupled to the $\left\vert g\right\rangle \leftrightarrow
\left\vert e\right\rangle $ transition of the coupler.}
\label{fig:3}
\end{figure}

\section{Implementation of an $n$-qubit MCP gate with photonic qubits}

Let us consider a system consisting of $n$\ cavities ($1,2,...,n$) coupled
to a two-level coupler (Fig.~3a). The coupler here is a SC qubit with two
levels $\left\vert g\right\rangle $\ and $\left\vert e\right\rangle $\
(Fig.~3b). Adjust the level spacings of the coupler such that cavity\ $j$\ ($%
j=1,2,...,n$)\quad is dispersively coupled to the $\left\vert g\right\rangle
$\ $\leftrightarrow $\ $\left\vert e\right\rangle $\ transition of the
coupler, with coupling constant $g_{j}$\ and detuning $\Delta _{j}$\
(Fig.~3b). Note that for the coupler being a SC qubit, the level spacings
can be rapidly (within $1-3$\ ns) adjusted through changing the external
control parameters (e.g., magnetic flux applied to the superconducting loop
of a SC phase, transmon [69], Xmon [70], or flux qubit [71]). In addition, a
classical pulse\ with frequency $\omega _{p}$\ and initial phase $\phi $\ is
applied to the coupler. The pulse is dispersively coupled to the $\left\vert
g\right\rangle $\ $\leftrightarrow $\ $\left\vert e\right\rangle $\
transition (Fig.~3b). In the interaction picture and after making a
rotating-wave approximation, the Hamiltonian of the system is given by
\begin{equation}
H=\sum\limits_{j=1}^{n}g_{j}e^{i\Delta _{j}t}\hat{a}_{j}\left\vert
e\right\rangle \left\langle g\right\vert +\Omega _{p}e^{i\left[ \left(
\omega _{eg}-\omega _{p}\right) t-\phi \right] }\left\vert e\right\rangle
\left\langle g\right\vert +\text{h.c.,}
\end{equation}%
where $\hat{a}_{j}$\ is the photon annihilation operator of cavity $j$, $%
\Omega _{p}$\ is the Rabi frequency of the pulse, and $\Delta _{j}=\omega
_{eg}-\omega _{c_{j}}$\ (Fig. 3b)$.$\ Here, $\omega _{eg}$\ is the $%
\left\vert g\right\rangle $\ $\leftrightarrow $\ $\left\vert e\right\rangle $
\ transition frequency of the coupler, while $\omega _{c_{j}}$\ is the
frequency of cavity $j$ ($j=1,2,...,n$).

For the dispersive coupling $\left\vert \Delta _{j}\right\vert \gg g_{j\text{
}}$(large detuning), the energy exchange between cavity $j$\ and the coupler can be neglected.
Under the condition $\left\vert \Delta _{j}-\Delta _{k}\right\vert
/\left( \left\vert \Delta _{j}^{-1}\right\vert +\left\vert \Delta
_{k}^{-1}\right\vert \right) \gg g_{j}g_{k}$\ (where $j,k\in
\{1,2,...n\},j\neq k$), the interaction between the cavities, induced
by the coupler, is negligible. In addition, assume $\left\vert \Delta _{j}\right\vert \gg
\Omega _{p}$ so that the stark-shift effect of the coupler induced by the
pulse is negligible.\ Under these considerations, based on the Hamiltonian
(3), one can obtain the following effective Hamiltonian [72,73]
\begin{equation}
H_{\mathrm{e}}=\sum\limits_{j=1}^{n}\lambda _{j}\left( \hat{a}_{j}^{+}\hat{a}%
_{j}+1/2\right) \sigma _{z}+\Omega _{p}e^{i\left[ \left( \omega _{eg}-\omega
_{p}\right) t-\phi \right] }\left\vert e\right\rangle \left\langle
g\right\vert +\text{h.c.},
\end{equation}%
where $\lambda _{j}=g_{j}^{2}/\Delta _{j}$\ and $\sigma _{z}=\left\vert
e\right\rangle \left\langle e\right\vert -\left\vert g\right\rangle
\left\langle g\right\vert .$\ In a rotating frame under the Hamiltonian $%
H_{0}=\sum\limits_{j=1}^{n}\lambda _{j}\left( \hat{a}_{j}^{+}\hat{a}%
_{j}+1/2\right) \sigma _{z}$\ and by choosing $\omega _{p}=\omega
_{eg}+\sum_{j=1}^{n}\lambda _{j}=\omega _{eg}+\sum_{j=1}^{n}g_{j}^{2}/\Delta
_{j},$\ it follows from the Hamiltonian (4)
\begin{equation}
H_{\mathrm{e}}=\Omega _{p}e^{-i\phi }e^{i2\sum_{j=1}^{n}\lambda _{j}\hat{a}%
_{j}^{+}\hat{a}_{j}t}\left\vert e\right\rangle \left\langle g\right\vert +%
\text{h.c.}
\end{equation}

For the gate purpose, we consider that each cavity is either in a vacuum
state $\left\vert 0\right\rangle $ or an arbitrary non-vacuum state
\begin{equation}
\left\vert \varphi \right\rangle =\sum\limits_{m=0}^{\infty }c_{m}\left\vert
m\right\rangle ,
\end{equation}%
where $\left\vert m\right\rangle $ is an $m$-photon Fock state. When the $n$
\ cavities are in the vacuum state $\left\vert 00...0\right\rangle ,$\ the
Hamiltonian (5) reduces to $H_{\mathrm{eff}}=\Omega _{p}e^{-i\phi
}\left\vert e\right\rangle \left\langle g\right\vert +$\ h.c., which rotates
the coupler's state as follows
\begin{equation}
\left\vert g\right\rangle \left\vert 00...0\right\rangle \rightarrow \left(
\cos \Omega _{p}t\left\vert g\right\rangle -ie^{-i\phi }\sin \Omega
_{p}t\left\vert e\right\rangle \right) \left\vert 00...0\right\rangle .
\end{equation}%
On the other hand, when the $n$\ cavities are not in the vacuum state, if
the Rabi frequency $\Omega _{p}$\ of the driving pulse is much smaller than $%
2\left\vert \lambda _{j}\right\vert \overline{n}_{j}$ (i.e., $\Omega _{p}\ll
2\left\vert \lambda _{j}\right\vert \overline{n}_{j}$)$,$\ the coupler's
state is not changed by the driving pulse due to the large detuning [74].
Here, $\overline{n}_{j}$ is the average photon number of cavity $j$ ($%
j=1,2,...,n$). In this sense, one has
\begin{equation}
\left\vert g\right\rangle \left\vert l_{1}l_{2}...l_{n}\right\rangle
\rightarrow \left\vert g\right\rangle \left\vert
l_{1}l_{2}...l_{n}\right\rangle ,
\end{equation}%
where subscript $j$ represents cavity $j$, $\left\vert
l_{1}l_{2}...l_{n}\right\rangle $\ is an abbreviation of the product state $%
\left\vert l_{1}\right\rangle $ $\left\vert l_{2}\right\rangle ...\left\vert
l_{n}\right\rangle $ of $n$\ cavities, and $\left\vert
l_{1}l_{2}...l_{n}\right\rangle \neq \left\vert 00...0\right\rangle $ (i.e.,
the $n$ cavities are not in the vacuum state). Here, $\left\vert
l_{j}\right\rangle $ $\in \left\{ \left\vert 0\right\rangle ,\left\vert
\varphi \right\rangle \right\} $ ($j=1,2,...,n$)$,$ which is the state of
cavity $j$.

By applying a unitary operation $U=e^{-iH_{0}t}$\ to return to the original
interaction picture, one has the following state transformation according to
Eqs.~(7) and (8)
\begin{equation}
\left\vert g\right\rangle \left\vert 00...0\right\rangle \rightarrow \left(
e^{i\sum_{j=1}^{n}\lambda _{j}t/2}\cos \Omega _{p}t\left\vert g\right\rangle
-ie^{-i\phi }e^{-i\sum_{j=1}^{n}\lambda _{j}t/2}\sin \Omega _{p}t\left\vert
e\right\rangle \right) \left\vert 00...0\right\rangle ,
\end{equation}
\begin{equation}
\left\vert g\right\rangle \left\vert l_{1}l_{2}...l_{n}\right\rangle
\rightarrow e^{i\sum_{j=1}^{n}\lambda _{j}t/2}e^{i\sum_{j=1}^{n}\lambda _{j}
\hat{a}_{j}^{+}\hat{a}_{j}t}\left\vert g\right\rangle \left\vert
l_{1}l_{2}...l_{n}\right\rangle .
\end{equation}
In the following, we set $\left\vert \lambda _{j}\right\vert =\lambda $\ ($%
j=1,2,...,n$). If the coupler-cavity interaction time is chosen such that $%
\lambda t=2k\pi $\ and $\Omega _{p}t=s\pi $\ ($k$\ is a positive integer
while $s$\ is a positive odd number), one has from Eqs.~(9) and (10)
\begin{eqnarray}
\left\vert g\right\rangle \left\vert 00...0\right\rangle &\rightarrow
&e^{i\sum_{j=1}^{n}\lambda _{j}t/2}\left( -\left\vert g\right\rangle
\left\vert 00...0\right\rangle \right) ,  \notag \\
\left\vert g\right\rangle \left\vert l_{1}l_{2}...l_{n}\right\rangle
&\rightarrow &e^{i\sum_{j=1}^{n}\lambda _{j}t/2}\left\vert g\right\rangle
e^{i\sum_{j=1}^{n}\pm 2k\pi \hat{a}_{j}^{+}\hat{a}_{j}}\left\vert
l_{1}l_{2}...l_{n}\right\rangle ,
\end{eqnarray}
where we take $2k\pi $ for $\lambda _{j}>0$ but $-2k\pi $ for $\lambda
_{j}<0.$ One can verify that the following equality holds
\begin{equation}
e^{i\sum_{j=1}^{n}\pm 2k\pi \hat{a}_{j}^{+}\hat{a}_{j}}\left\vert
l_{1}l_{2}...l_{n}\right\rangle =\left\vert l_{1}l_{2}...l_{n}\right\rangle .
\end{equation}

To see Eq.~(12) clearly, let us consider a three-cavity case (i.e., $n=3$).
Because of $\left\vert l_{j}\right\rangle $ $\in \left\{ \left\vert
0\right\rangle ,\left\vert \varphi \right\rangle \right\} $ ($j=1,2,3$), the
product state $\left\vert l_{1}l_{2}l_{3}\right\rangle $ of the three
cavities is $\left\vert 00\varphi \right\rangle ,\left\vert 0\varphi
0\right\rangle$, $\left\vert 0\varphi \varphi \right\rangle$, $\left\vert
\varphi 00\right\rangle$, $\left\vert \varphi 0\varphi \right\rangle$, $%
\left\vert \varphi \varphi 0\right\rangle$, or $\left\vert \varphi \varphi
\varphi \right\rangle$. By applying $e^{i\sum_{j=1}^{3}\pm 2k\pi \hat{a}
_{j}^{+}\hat{a}_{j}}$ to these states, we have
\begin{eqnarray}
e^{i\sum_{j=1}^{3}\pm 2k\pi \hat{a}_{j}^{+}\hat{a}_{j}}\left\vert 00\varphi
\right\rangle &=&\left\vert 00\right\rangle \sum\limits_{m=0}^{\infty
}c_{m}e^{\pm i2mk\pi }\left\vert m\right\rangle =\left\vert 00\varphi
\right\rangle ,  \notag \\
e^{i\sum_{j=1}^{3}\pm 2k\pi \hat{a}_{j}^{+}\hat{a}_{j}}\left\vert 0\varphi
0\right\rangle &=&\left\vert 0\right\rangle \sum\limits_{m=0}^{\infty
}c_{m}e^{\pm i2mk\pi }\left\vert m\right\rangle \left\vert 0\right\rangle
=\left\vert 0\varphi 0\right\rangle ,  \notag \\
e^{i\sum_{j=1}^{3}\pm 2k\pi \hat{a}_{j}^{+}\hat{a}_{j}}\left\vert 0\varphi
\varphi \right\rangle &=&\left\vert 0\right\rangle\sum\limits_{m=0}^{\infty
}c_{m}e^{\pm i2mk\pi }\left\vert m\right\rangle \sum\limits_{m=0}^{\infty
}c_{m}e^{\pm i2mk\pi }\left\vert m\right\rangle =\left\vert 0\varphi \varphi
\right\rangle ,  \notag \\
e^{i\sum_{j=1}^{3}\pm 2k\pi \hat{a}_{j}^{+}\hat{a}_{j}}\left\vert \varphi
00\right\rangle &=&\sum\limits_{m=0}^{\infty }c_{m}e^{\pm i2mk\pi
}\left\vert m\right\rangle \left\vert 0\right\rangle \left\vert
0\right\rangle =\left\vert \varphi 00\right\rangle ,  \notag \\
e^{i\sum_{j=1}^{3}\pm 2k\pi \hat{a}_{j}^{+}\hat{a}_{j}}\left\vert \varphi
0\varphi \right\rangle &=&\sum\limits_{m=0}^{\infty }c_{m}e^{\pm i2mk\pi
}\left\vert m\right\rangle \left\vert 0\right\rangle
\sum\limits_{m=0}^{\infty }c_{m}e^{\pm i2mk\pi }\left\vert m\right\rangle
=\left\vert \varphi 0\varphi \right\rangle ,  \notag \\
e^{i\sum_{j=1}^{3}\pm 2k\pi \hat{a}_{j}^{+}\hat{a}_{j}}\left\vert \varphi
\varphi 0\right\rangle &=&\sum\limits_{m=0}^{\infty }c_{m}e^{\pm i2mk\pi
}\left\vert m\right\rangle \sum\limits_{m=0}^{\infty }c_{m}e^{\pm i2mk\pi
}\left\vert m\right\rangle \left\vert 0\right\rangle =\left\vert \varphi
\varphi 0\right\rangle ,  \notag \\
e^{i\sum_{j=1}^{3}\pm 2k\pi \hat{a}_{j}^{+}\hat{a}_{j}}\left\vert \varphi
\varphi \varphi \right\rangle &=&\sum\limits_{m=0}^{\infty }c_{m}e^{\pm
i2mk\pi }\left\vert m\right\rangle \sum\limits_{m=0}^{\infty }c_{m}e^{\pm
i2mk\pi }\left\vert m\right\rangle \sum\limits_{m=0}^{\infty }c_{m}e^{\pm
i2mk\pi }\left\vert m\right\rangle  \notag \\
&=&\left\vert \varphi \varphi \varphi \right\rangle ,
\end{eqnarray}
where we have used $\left\vert \varphi \right\rangle
=\sum\limits_{m=0}^{\infty }c_{m}\left\vert m\right\rangle $ (see Eq.~(6))
and applied $e^{\pm i2mk\pi }=1.$ Obviously, the state transformations given
in Eq.~(13) can be summarized as Eq.~(12) for $n=3.$ Following the same
derivation as shown in Eq. (13), one can easily verify that Eq.~(12) also
holds for the case of $n>3.$

According to Eq.~(12), we have from Eq.~(11)
\begin{eqnarray}
\left\vert g\right\rangle \left\vert 00...0\right\rangle &\rightarrow
&e^{i\sum_{j=1}^{n}\lambda _{j}t/2}\left( -\left\vert g\right\rangle
\left\vert 00...0\right\rangle \right) ,  \notag \\
\left\vert g\right\rangle \left\vert l_{1}l_{2}...l_{n}\right\rangle
&\rightarrow &e^{i\sum_{j=1}^{n}\lambda _{j}t/2}\left\vert g\right\rangle
\left\vert l_{1}l_{2}...l_{n}\right\rangle ,
\end{eqnarray}
After dropping off the common phase factor $e^{i\sum_{j=1}^{n}\lambda
_{j}t/2},$\ it follows from Eq. (14)
\begin{eqnarray}
\left\vert 00...0\right\rangle \left\vert g\right\rangle &\rightarrow
&-\left\vert 00...0\right\rangle \left\vert g\right\rangle ,  \notag \\
\left\vert l_{1}l_{2}...l_{n}\right\rangle \left\vert g\right\rangle
&\rightarrow &\left\vert l_{1}l_{2}...l_{n}\right\rangle \left\vert
g\right\rangle ,\text{ }
\end{eqnarray}
where $\left\vert l_{1}l_{2}...l_{n}\right\rangle \neq \left\vert
00...0\right\rangle .$

Let us now consider $n$ photonic qubits ($1,2,...,n$). Each photonic qubit
is encoded as follows. Namely, the logic state $\left\vert 1\right\rangle
_{L}$\ of photonic qubit $j$ is represented by the vacuum state $\left\vert
0\right\rangle $\ of cavity $j,$ while the logic state $\left\vert
0\right\rangle _{L}$\ of photonic qubit $j$ is represented by the non-vacuum
state $\left\vert \varphi \right\rangle $ of cavity $j$ ($j=1,2,...,n$).
With this encoding, the state $\left\vert 00...0\right\rangle $ of the $n$
cavities corresponds to the computational basis state $\left\vert
11...1\right\rangle $ of the $n$ photonic qubits, and the state $\left\vert
l_{1}l_{2}...l_{n}\right\rangle $ of the $n$ cavities corresponds to the
computational basis state $\left\vert i_{1}i_{2}...i_{n}\right\rangle $ of
the $n$ photonic qubits ($i_{j}\in \left\{ 0,1\right\} $). Note that we have
$\left\vert i_{1}i_{2}...i_{n}\right\rangle \neq \left\vert
11...1\right\rangle $ because of $\left\vert l_{1}l_{2}...l_{n}\right\rangle
\neq \left\vert 00...0\right\rangle $ (see the above). Thus, from the state
transformation of the $n$ cavities in Eq.~(15), one can obtain the following
state transformation for the $n$ photonic qubits
\begin{eqnarray}
\left\vert 11...1\right\rangle &\rightarrow &-\left\vert 11...1\right\rangle
,  \notag \\
\left\vert i_{1}i_{2}...i_{n}\right\rangle &\rightarrow &\left\vert
i_{1}i_{2}...i_{n}\right\rangle ,\text{ for }\left\vert
i_{1}i_{2}...i_{n}\right\rangle \neq \left\vert 11...1\right\rangle .
\end{eqnarray}
where subscripts ($1,2,...,n$) represent the $n$ photonic qubits $1,2,...,$
and $n$, respectively. Note that the state transformation (16) is identical
to that given in Eq. (1). Hence, an $n$-qubit MCP gate described by Eq. (1)
is implemented after the above operation. In addition, one can see from Eq.
(15) that the coupler returns to its original ground state after the
operation. \

Before ending this section, some points may need to be addressed here:

(a) In above we have set $\left\vert \lambda _{j}\right\vert =\lambda $\ ($%
j=1,2,...,n$), which turns out into
\begin{equation}
g_{1}^{2}/\left\vert \Delta _{1}\right\vert =g_{2}^{2}/\left\vert \Delta
_{2}\right\vert =...=g_{n}^{2}/\left\vert \Delta _{n}\right\vert .
\end{equation}
This condition can be readily achieved by carefully selecting the detunings $%
\Delta _{1},\Delta _{2},...,\Delta _{n}$\ via prior adjustment of the cavity
frequencies due to $\Delta _{j}=\omega _{eg}-\omega _{c_{j}}$\ ($j=1,2,...,n$
).

(b) In above we have set
\begin{equation}
\lambda t=2k\pi ,\text{ }\Omega _{p}t=s\pi ,
\end{equation}
which results in
\begin{equation}
\Omega _{p}=\frac{s}{2k}\lambda .
\end{equation}
The condition (19)\ can be easily satisfied by adjusting the Rabi frequency $%
\Omega _{p}$\ of the pulse (e.g., through varying the pulse intensity).

(c) As mentioned above, the Hamiltonian (3) was constructed by adjusting the
level spacings of the coupler. However, we should point out that adjusting
the level spacings of the coupler is unnecessary. Alternatively, one can
obtain the Hamiltonian (3) by tuning the frequency of each cavity. Note that
the frequency of a superconducting microwave cavity or resonator can be
rapidly tuned within a few nanoseconds [75,76].

(d) The single-step implementation of the $n$-qubit MCP gate here can
significantly simplify the realization of an $n$-qubit Toffoli gate (Fig.
1a). This is because the $n$-qubit Toffoli gate can be constructed by
combining the $n$-qubit MCP gate with two single-qubit Hadamard gates
[13,28], which are performed on the target qubit before and after the $n$
-qubit MCP gate respectively (Fig.1c). However, using the conventional
gate-constructing technique to construct a Toffoli gate, the required number
of single- and two-qubit quantum gates increases drastically with an
increasing number of qubits [9-12].

(e) As shown above, the $n$-qubit MCP gate is realized through a single step
of operation essentially described by the effective Hamiltonian (5), which
is derived from the original Hamiltonian (3). In addition, neither
measurement on the state of cavities nor measurement on the state of the
coupler is needed.

\section{Encoding and orthogonality}

In the above, the two logic states of each photonic qubit are encoded with a
vacuum state $\left\vert 0\right\rangle $ and an arbitrary non-vacuum state $%
\left\vert \varphi \right\rangle $. We now give a brief discussion on the
orthogonality required by the qubit encoding. In other words, to have the
encoding effective, the orthogonality or quasi-orthogonality between the
vacuum state $\left\vert 0\right\rangle $ and the non-vacuum state $%
\left\vert \varphi \right\rangle $ needs to hold, i.e.,
\begin{equation}
\left\langle 0\right\vert \left. \varphi \right\rangle \simeq 0.
\end{equation}
To name a few, we will provide some encodings for which the condition (20)
applies:

(i) The two logic states of each photonic qubit are encoded with a vacuum
state $\left\vert 0\right\rangle $ and a Fock state $\left\vert
m\right\rangle $ with $m$ photons (i.e., $\left\vert \varphi \right\rangle
=\left\vert m\right\rangle $). For this encoding, one has $\left\langle
0\right\vert \left. \varphi \right\rangle =0.$

(ii) The two logic states of each photonic qubit are encoded with a vacuum
state $\left\vert 0\right\rangle $ and a superposition of Fock states (e.g.,
$\left\vert \varphi \right\rangle =\frac{1}{\sqrt{2}}\left( \left\vert
1\right\rangle +\left\vert 2\right\rangle \right) ,\frac{1}{\sqrt{n}}\left(
\left\vert 1\right\rangle +\left\vert 2\right\rangle +...+\left\vert
n\right\rangle \right) ,$ etc.). For this encoding, one has $\left\langle
0\right\vert \left. \varphi \right\rangle =0.$

(iii) The two logic states of each photonic qubit are encoded with a vacuum
state $\left\vert 0\right\rangle $ and a Schr\"{o}dinger cat state (e.g., $%
\left\vert \varphi \right\rangle =\mathcal{N}\left( \left\vert \alpha
\right\rangle -\left\vert -\alpha \right\rangle \right) ,$ where $\left\vert
\pm \alpha \right\rangle $ are coherent states and $\mathcal{N}$ is a
normalization factor). For this encoding, one has $\left\langle 0\right\vert
\left. \varphi \right\rangle =0.$ It is noted that the cat-state encoding,
consisting of superpositions of coherent states, is protected against photon
loss and dephasing errors [77,78], and quantum computing based on cat-state
encoding has recently attracted much attention [68,79--81].

(iv)\ The two logic states of each photonic qubit are encoded with a vacuum
state $\left\vert 0\right\rangle $ and a coherent state $\left\vert \alpha
\right\rangle $ with a large enough $\alpha .$ One can check $\left\langle
0\right\vert \left. \alpha \right\rangle =\exp \left( -\left\vert \alpha
\right\vert ^{2}/2\right) \approx 0$ for a large enough $\alpha .$

(v) The two logic states of each photonic qubit are encoded with a vacuum
state $\left\vert 0\right\rangle $ and a multi-component Schr\"{o}dinger cat
state (e.g., $\left\vert \varphi \right\rangle =\mathcal{N}\left( \left\vert
\alpha \right\rangle -\left\vert -\alpha \right\rangle +\left\vert i\alpha
\right\rangle -\left\vert -i\alpha \right\rangle \right) ,$ where $%
\left\vert \pm \alpha \right\rangle $ and $\left\vert \pm i\alpha
\right\rangle $ are coherent states and $\mathcal{N}$ is a normalization
factor). For this encoding, one has $\left\langle 0\right\vert \left.
\varphi \right\rangle =0.$

(vi) The two logic states of each photonic qubit are encoded with a vacuum
state $\left\vert 0\right\rangle $ and a squeezed vacuum state $\left\vert
\xi \right\rangle $ with a large enough squeezed parameter $r.$ Here, $\xi
=re^{i\theta }.$ One can verify $\left\langle 0\right\vert \left. \xi
\right\rangle =\sqrt{2/\left( e^{r}+e^{-r}\right) }\approx 0$ for a large
enough $r.$

\begin{figure}[tbp]
\begin{center}
\includegraphics[bb=285 265 482 446, width=6.5 cm, clip]{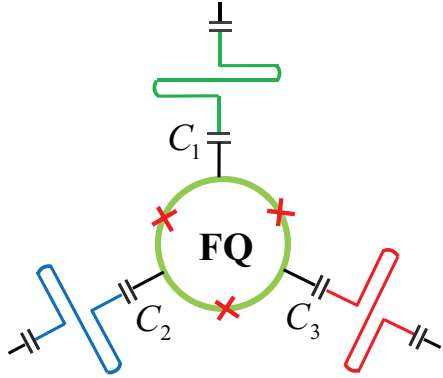} \vspace*{%
				-0.08in}
\end{center}
\caption{(color online) Diagram of three 1D microwave cavities capacitively
coupled to a superconducting flux qubit (FQ). Each cavity here is a
one-dimensional transmission line resonator. The flux qubit consists of
three Josephson junctions and a superconducting loop.}
\label{fig:4}
\end{figure}

\begin{figure}[tbp]
\begin{center}
\includegraphics[bb=487 275 675 491, width=6.5 cm, clip]{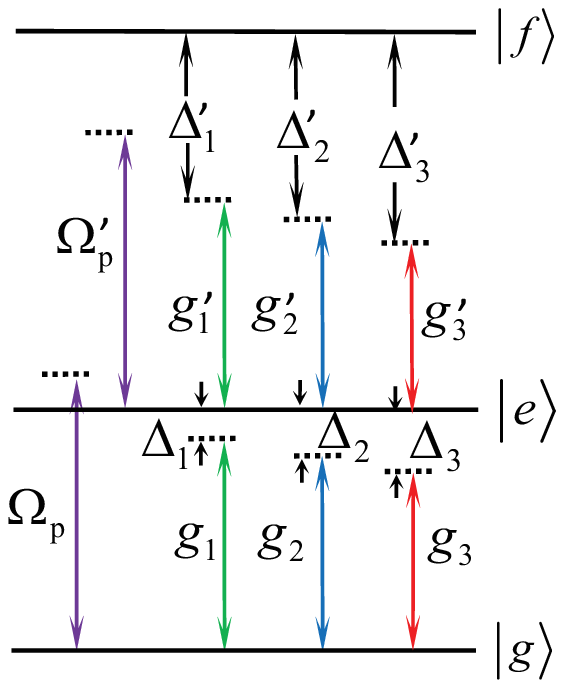} \vspace*{%
				-0.08in}
\end{center}
\caption{(color online) Illustration of three cavities (1,2,3) and a
microwave pulse dispersively coupled to the $\left\vert g\right\rangle
\leftrightarrow \left\vert e\right\rangle $ transition of the flux qubit, as
well as the unwanted couplings of the three cavities and the pulse with the $%
\left\vert e\right\rangle \leftrightarrow \left\vert f\right\rangle $
transition of the flux qubit. For a flux qubit, the level spacing between
the upper two levels is larger than that between the two lowest levels.}
\label{fig:5}
\end{figure}

\section{Possible experimental feasibility}

In the above, we have considered a two-level SC qubit as the coupler to
couple multiple cavities. The SC qubit could be a SC phase, flux, transmon,
or Xmon qubit, etc. As an example, we now give a discussion on the
experimental feasibility of realizing a three-qubit MCP gate based on a
circuit-QED system, which consists of three 1D microwave cavities (1,2,3)
coupled to a SC flux qubit (Fig.~4). In this example, we consider that the
two logic states $\left\vert 0\right\rangle _{L}$ and $\left\vert
1\right\rangle _{L}$ of each photonic qubit are encoded with a cat state $%
\mathcal{N}\left( \left\vert \alpha \right\rangle -\left\vert -\alpha
\right\rangle \right) $ and a vacuum state. In this case, we have
\begin{equation}
\left\vert \varphi \right\rangle =\mathcal{N}\left( \left\vert \alpha
\right\rangle -\left\vert -\alpha \right\rangle \right) ,
\end{equation}
with $\mathcal{N=}1/\sqrt{2\left( 1+e^{-2\left\vert \alpha \right\vert
2}\right) }.$ The three photonic qubits (1,2,3) involved in the gate
correspond to three microwave cavities (1,2,3), respectively.

In reality, we need to consider the effect of the second excited level $%
\left\vert f\right\rangle $ of the flux qubit and the inter-cavity crosstalk
on the gate operation. Thus, we modify the Hamiltonian (3) as $H^{\prime
}=H+\delta H.$ Here, $H$ is the Hamiltonian (3) given above (with $n=3$),
and $\delta H$ is given by
\begin{eqnarray}
\delta H &=&\left( \sum\limits_{j=1}^{3}g_{j}^{\prime }e^{i\Delta
_{j}^{\prime }t}\hat{a}_{j}\left\vert f\right\rangle \left\langle
e\right\vert +\text{h.c.}\right) +\left( \Omega _{p}^{\prime }e^{i\left[
\left( \omega _{fe}-\omega _{p}\right) t-\phi \right] }\left\vert
f\right\rangle \left\langle e\right\vert +\text{h.c.}\right)  \notag \\
&&+\left( g_{12}e^{i\Delta _{12}}\hat{a}_{1}^{+}\hat{a}_{2}+g_{23}e^{i\Delta
_{23}}\hat{a}_{2}^{+}\hat{a}_{3}+g_{13}e^{i\Delta _{13}}\hat{a}_{1}^{+}\hat{%
a }_{3}+\text{h.c.}\right) \text{,}
\end{eqnarray}
where the terms in the first bracket represent the unwanted coupling of the
three cavities with the $\left\vert e\right\rangle \leftrightarrow
\left\vert f\right\rangle $ transition with coupling constant $g_{j}^{\prime
}$ and detuning $\Delta _{j}^{\prime }=\omega _{fe}-\omega _{c_{j}}$ ($%
j=1,2,3$) (Fig.~5), the terms in the second bracket represent the unwanted
coupling between the pulse and the $\left\vert e\right\rangle
\leftrightarrow \left\vert f\right\rangle $ transition with Rabi frequency $%
\Omega _{p}^{\prime }$ (Fig.~5), while the terms in the last bracket
represent the inter-cavity crosstalk with the crosstalk strength $g_{kl}$
and frequency detuning $\Delta _{kl}=\omega _{c_{k}}-\omega _{c_{l}}$
between the two cavities $k$ and $l.$ Here, $kl\in \left\{ 12,23,13\right\}
. $ Note that the coupling of the cavities and the pulse with the $%
\left\vert g\right\rangle \leftrightarrow \left\vert f\right\rangle $
transition is negligible because of $\omega _{eg},\omega _{fe}\ll \omega
_{fg}$ (Fig.~5). Here, $\omega _{fe}$ ($\omega _{fg}$) is the $\left\vert
e\right\rangle \leftrightarrow \left\vert f\right\rangle $ ($\left\vert
g\right\rangle \leftrightarrow \left\vert f\right\rangle $) transition
frequency of the qubit.

After considering the system dissipation and dephasing, the dynamics of the
lossy system is determined by the master equation
\begin{eqnarray}
\frac{d\rho }{dt} &=&-i\left[ H^{\prime },\rho \right] +\sum_{j=1}^{3}\kappa
_{j}\mathcal{L}\left[ \hat{a}_{j}\right]  \notag \\
&&+\gamma _{eg}\mathcal{L}\left[ \sigma _{eg}^{-}\right] +\gamma _{fe}
\mathcal{L}\left[ \sigma _{fe}^{-}\right] +\gamma _{fg}\mathcal{L}\left[
\sigma _{fg}^{-}\right]  \notag \\
&&+\gamma _{e,\varphi }\left( \sigma _{ee}\rho \sigma _{ee}-\sigma _{ee}\rho
/2-\rho \sigma _{ee}/2\right)  \notag \\
&&+\gamma _{f,\varphi }\left( \sigma _{ff}\rho \sigma _{ff}-\sigma _{ff}\rho
/2-\rho \sigma _{ff}/2\right) ,
\end{eqnarray}
where $\rho $ is the density matrix of the whole system; $H^{\prime }$ is
the modified Hamiltonian given above; $\mathcal{L}\left[ \xi \right] =\xi
\rho \xi ^{+}-\xi ^{+}\xi \rho /2-\rho \xi ^{+}\xi /2$\ (with $\xi =\hat{a}
_{j},\sigma _{eg}^{-},\sigma _{fe}^{-},\sigma _{fg}^{-}$),\ $\sigma
_{eg}^{-}=\left\vert g\right\rangle \left\langle e\right\vert ,$\ $\sigma
_{fe}^{-}=\left\vert e\right\rangle \left\langle f\right\vert ,$\ $\sigma
_{fg}^{-}=\left\vert g\right\rangle \left\langle f\right\vert ,$\ $\sigma
_{ee}=\left\vert e\right\rangle \left\langle e\right\vert $, and $\sigma
_{ff}=\left\vert f\right\rangle \left\langle f\right\vert ;$\ $\gamma
_{e,\varphi }$\ ($\gamma _{f,\varphi }$) is the dephasing rate of the level $%
\left\vert e\right\rangle $\ ($\left\vert f\right\rangle $); $\gamma _{eg}$\
is the $\left\vert e\right\rangle \rightarrow \left\vert g\right\rangle $
energy relaxation rate of the level $\left\vert e\right\rangle $; $\gamma
_{fe}$\ ($\gamma _{fg}$) is the $\left\vert f\right\rangle \rightarrow
\left\vert e\right\rangle $\ ($\left\vert f\right\rangle \rightarrow
\left\vert g\right\rangle $) energy relaxation rate of the level $\left\vert
f\right\rangle $ of the qubit; while $\kappa _{j}$\ is the decay rate of
cavity $j$\ ($j=1,2,3$). \

The fidelity of the whole operation is given by $F=\sqrt{\left\langle \psi
_{ \mathrm{id}}\right\vert \rho \left\vert \psi _{\mathrm{id}}\right\rangle }%
,$ where $\left\vert \psi _{\mathrm{id}}\right\rangle $ is the ideal output
state obtained under the theoretical model, while $\rho $ is the final
density matrix of the whole system (i.e., the three cavities and the flux
qubit) obtained by numerically solving the master equation. As an example,
let us consider an input state of the whole system $\frac{1}{2\sqrt{2}}\sum
\left\vert l_{1}l_{2}l_{3}\right\rangle \otimes \left\vert g\right\rangle ,$
where $\left\vert l_{j}\right\rangle \in \left\{ \left\vert 0\right\rangle
,\left\vert \varphi \right\rangle \right\} $ ($j=1,2,3$)$,$ with $\left\vert
\varphi \right\rangle $ given in Eq. (21). Thus, according to Eq. (15), the
ideal output state of the whole system is $\left\vert \psi _{\mathrm{id}
}\right\rangle =\frac{1}{2\sqrt{2}}\left( \sum_{l_{1},l_{2},l_{3}\neq
0}\left\vert l_{1}l_{2}l_{3}\right\rangle -\left\vert 000\right\rangle
\right) \otimes \left\vert g\right\rangle $ after applying a three-qubit MCP
gate.

\begin{table}[tbp]
\centering
\begin{tabular*}{13.5cm}{lll}
\hline\hline
\ \ $\omega_{eg}/2\pi=6.5 \mathrm{\,GHz}$ \ \ \ \ \ \ \ \ \ \ $%
\omega_{fe}/2\pi=13.5 \mathrm{\,GHz}$ \,\ \ \ \ \ \ \ \ \ \ \ $%
\omega_{fg}/2\pi=20.0 \mathrm{\,GHz}$ &  &  \\[1.5ex]
\ \ \,\!$\omega_{c_{1}}/2\pi=4.5 \mathrm{\,GHz}$ \ \ \,\ \ \ \ \ \ \ \ $%
\omega_{c_{2}}/2\pi=3.72 \mathrm{\,GHz}$ \ \ \ \ \ \ \ \ \ \ \ $%
\omega_{c_{3}}/2\pi= 3.0 \mathrm{\,GHz}$ &  &  \\[1.5ex]
\ \ \ \!$\triangle_{1}/2\pi= 2.0 \mathrm{\,GHz}$ \quad \ \ \ \ \ \ \ \, $%
\triangle_{2}/2\pi= 2.78 \mathrm{\,GHz}$ \ \,\ \ \ \ \ \ \ \ \ \,$%
\triangle_{3}/2\pi= 3.5 \mathrm{\,GHz}$ &  &  \\[1.5ex]
\ \ \ \ \!\!\!$\triangle^{\prime}_{1}/2\pi= 9.0 \mathrm{\,GHz}$ \ \ \ \ \ \
\ \ \ \ \ \!$\triangle^{\prime}_{2}/2\pi= 9.78 \mathrm{\,GHz}$ \ \,\,\ \ \ \
\ \ \ \ \ \ \!$\triangle^{\prime}_{3}/2\pi= 10.5 \mathrm{\,GHz}$ &  &  \\%
[1.5ex]
\ \, $\triangle_{12}/2\pi= 0.78 \mathrm{\,GHz}$ \ \;\ \ \ \ \ \ \,$%
\triangle_{23}/2\pi= 0.72 \mathrm{\,GHz}$ \;\;\,\, \ \ \ \ \ \ \ \!$%
\triangle_{13}/2\pi= 1.5 \mathrm{\,GHz}$ &  &  \\[1.5ex]
\!\!\!\!\!\!\!\!\quad \ \,\,\,\, $g_{1}/2\pi= 0.123 \mathrm{\,GHz}$ \ \ \ \
\ \ \ \,\ \ \,\!$g_{2}/2\pi= 0.145 \mathrm{\,GHz}$ \quad \ \ \ \ \ \ \ \ \
\!\!$g_{3}/2\pi= 0.163 \mathrm{\,GHz}$ &  &  \\[1.5ex]
\!\!\!\!\!\!\!\!\quad \ \,\,\,\, $g_{1}^{\prime}/2\pi= 0.123 \mathrm{\,GHz}$
\ \ \ \ \ \ \ \,\ \ \,\!$g_{2}^{\prime}/2\pi= 0.145 \mathrm{\,GHz}$ \quad \
\ \ \ \ \ \ \ \ \!\!$g_{3}^{\prime}/2\pi= 0.163 \mathrm{\,GHz}$ &  &  \\%
[1.5ex]
\ \ \ $g_{12}/2\pi=1.63 \mathrm{\,MHz}$ \quad \ \ \ \ \ \ \ \!$%
g_{23}/2\pi=1.63 \mathrm{\,MHz}$ \quad \ \ \ \ \ \ \ \ $g_{13}/2\pi=1.63
\mathrm{\,MHz}$ &  &  \\[1.5ex]
\ \,\,\, $\Omega_{p}/2\pi=1.89 \mathrm{\,MHz}$ \quad \ \ \ \ \ \ $%
\Omega_{p}^{\prime}/2\pi=1.89 \mathrm{\,MHz}$ \quad \ \ \ \ \ \ \ \ \ $%
\omega_p/2\pi=6.523 \mathrm{\,GHz}$ &  &  \\ \hline\hline
\end{tabular*}%
\caption{Parameters used in the numerical simulation. $\protect\omega _{eg},$
$\protect\omega _{fe},$ and $\protect\omega _{fg}$ are the $\left\vert
g\right\rangle \leftrightarrow \left\vert e\right\rangle ,$ $\left\vert
e\right\rangle \leftrightarrow \left\vert f\right\rangle ,$ and $\left\vert
g\right\rangle \leftrightarrow \left\vert f\right\rangle $ transition
frequencies of the flux qubit, respectively. $\protect\omega _{c_{j}}$ is
the frequency of cavity $j$ ($j=1,2,3$). $\Delta _{j}$ ($\Delta
_{j}^{\prime} $) is the detuning between the frequency of cavity $j$ and the
$\left\vert g\right\rangle \leftrightarrow \left\vert e\right\rangle $ ($%
\left\vert e\right\rangle \leftrightarrow \left\vert f\right\rangle $)
transition frequency of the flux qubit ($j=1,2,3$). $\Delta _{kl}$ is the
frequency detuning between the two cavities $k$ and $l$ ($kl=12,23,13$). $%
g_{j}$ ($g_{j}^{\prime }$) is the coupling constant between cavity $j$ and
the $\left\vert g\right\rangle \leftrightarrow \left\vert e\right\rangle $ $%
\left( \left\vert e\right\rangle \leftrightarrow \left\vert f\right\rangle
\right) $ transition of the flux qubit. $g_{kl}$ is the crosstalk strength
between the two cavities $k$ and $l$ ($kl=12,23,13$). $\Omega _{p}$ ($\Omega
_{p}^{\prime }$) is the Rabi frequency of the pulse associated with the $%
\left\vert g\right\rangle \leftrightarrow \left\vert e\right\rangle $ ($%
\left\vert e\right\rangle \leftrightarrow \left\vert f\right\rangle $)
transition of the flux qubit. $\protect\omega_p$ is the frequency of the
pulse.}
\end{table}

For a superconducting flux qubit, the transition frequency between adjacent
energy levels can be 1 to 20 GHz [82-84]. As an example, consider the
parameters listed in Table 1, which are used in our numerical simulations.
The coupling constants $g_{2}$ and $g_{3}$ are calculated according to Eq.
(17). In addition, $\Omega _{p}$ is calculated for $s=1$ and $k=2$ according
to Eq.~(19). The pulse frequency $\omega _{p}$\ is calculated based on $%
\omega _{p}=\omega _{eg}+\sum_{j=1}^{3}g_{j}^{2}/\Delta _{j}.$ The dipole
matrix elements between any two of the three levels $\left\vert
g\right\rangle $, $\left\vert e\right\rangle $, and $\left\vert
f\right\rangle $ can be made to be on the same order of magnitude via
properly designing the flux qubit [85]. Thus, we choose $\Omega _{p}^{\prime
}\sim \Omega _{p}$ and $g_{j}^{\prime }\sim g_{j}$ ($j=1,2,3$) for
simplicity. In addition, we consider $g_{kl}=0.01g_{\max }$, with $g_{\max
}=\max \{g_{1},g_{2},g_{3}\}$ ($kl=12,23,13$). Other parameters used in the
numerical simulations are: (i) $\gamma _{eg}^{-1}=$ $2T$, $\gamma
_{fe}^{-1}=2T$, $\gamma _{fg}^{-1}=T$, (ii) $\gamma _{e,\varphi
}^{-1}=\gamma _{f,\varphi }^{-1}=T$, (iii) $\kappa _{j}=\kappa $ ($j=1,2,3$
), (iv) $\alpha =1.0$. 

\begin{figure}[tbp]
\begin{center}
\includegraphics[bb=0 0 541 354, width=11.5 cm, clip]{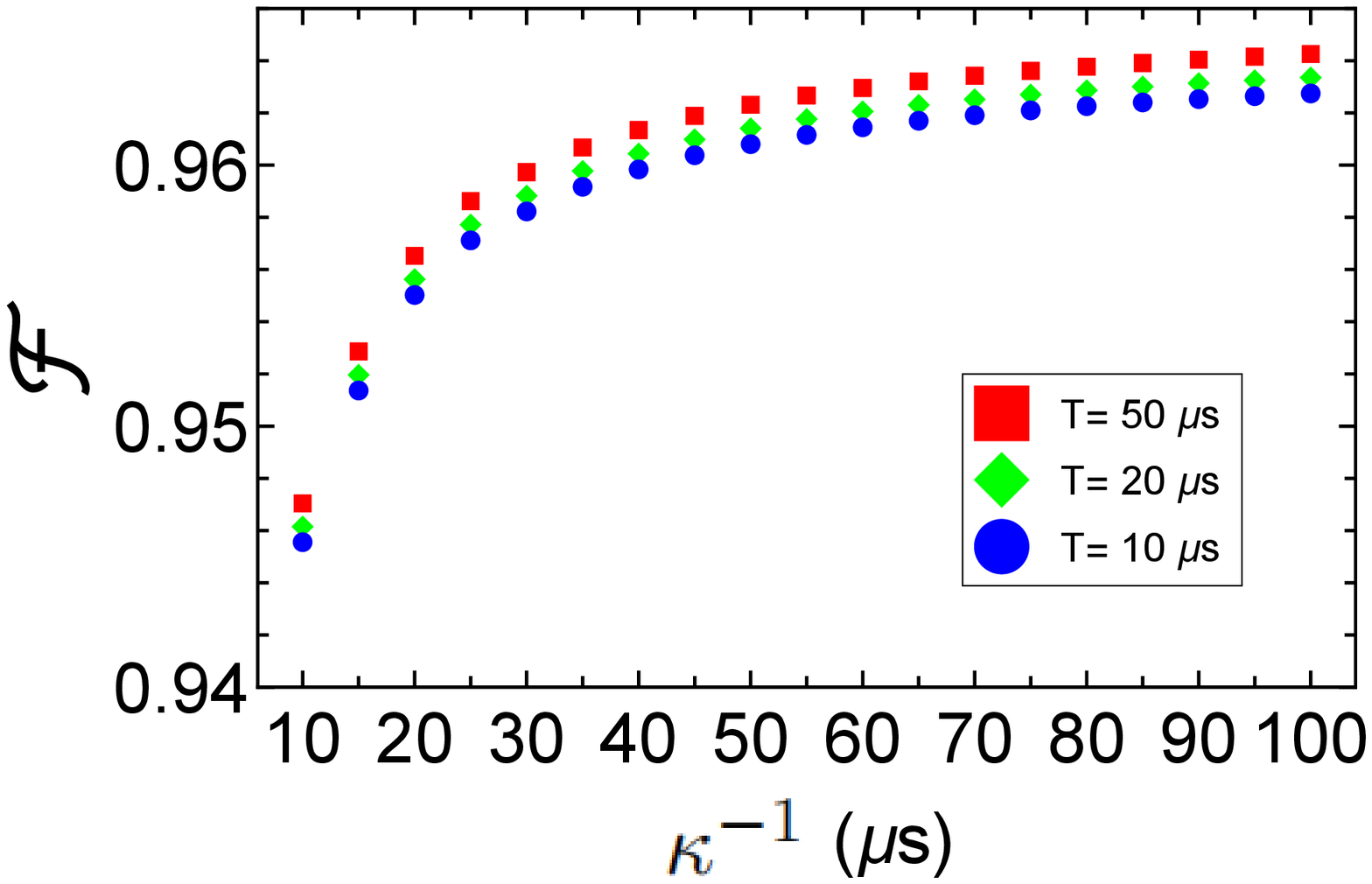} \vspace*{%
				-0.08in}
\end{center}
\caption{(Color online) Fidelity versus $\protect\kappa ^{-1}.$ Other
parameters used in the numerical simulation are referred to the text and
Table~1.}
\label{fig:6}
\end{figure}

\begin{figure}[tbp]
\begin{center}
\includegraphics[bb=0 0 539 352, width=11.5 cm, clip]{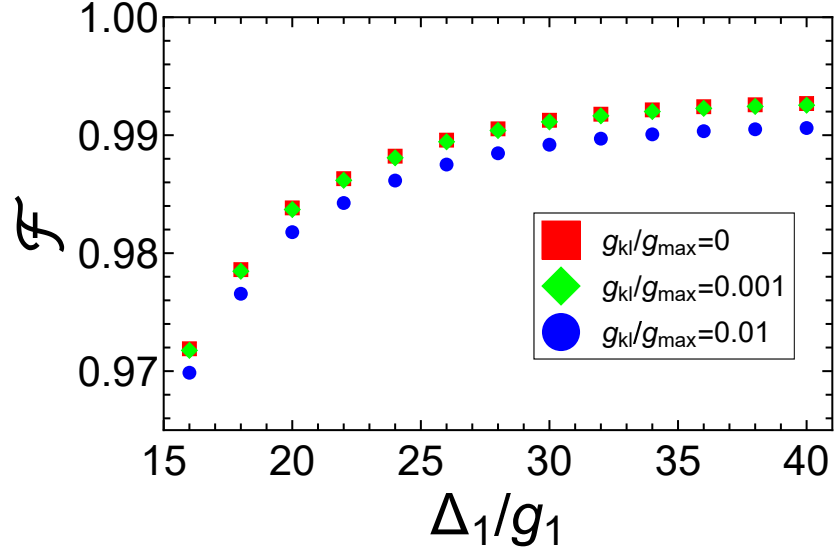} \vspace*{%
				-0.08in}
\end{center}
\caption{(Color online) Fidelity versus $\Delta _{1}/g_{1}$. The figure is
plotted by setting $\Delta _{2}=\Delta _{1}+10g_{1}, $ $\Delta _{3}=\Delta
_{1}+20g_{1},$ $\Omega _{p}=\frac{s}{2k}g_{1}^{2}/\Delta_{1},$ $g_{2}=
\protect\sqrt{\Delta _{2}/\Delta _{1}}g_{1},$ $g_{3}=\protect\sqrt{ \Delta
_{3}/\Delta _{1}}g_{1},$ and assuming that the dissipation of the system as
well as the unwanted couplings between the qubit levels $\left\vert
e\right\rangle $ and $\left\vert f\right\rangle $ (see Fig. 6) are
negligible.}
\label{fig:7}
\end{figure}

By solving the master equation (23), we numerically plot Fig. 6 to
illustrate the fidelity versus $\kappa ^{-1}$ for $T=10,20,50$ $\mu $s.
Figure 6 shows that when the decay rate of the cavities increases, the gate
fidelity quickly drops. This is because the present work focuses on the gate
with photonic qubits and thus photons are always populated in each cavity
during the gate realization. Nevertheless, one can see from Fig. 6 that for $%
\kappa ^{-1}$\ $\geq 35$\ $\mu $s and $T$\ $\geq 20$\ $\mu $s, the fidelity
exceeds $96\%$. The imperfect fidelity is also caused by the unwanted
couplings of the cavities (and the pulse) with the irrelevant levels of the
flux qubit as well as the decoherence of the flux qubit. In addition, the
imperfect fidelity is caused due to that the large detuning conditions are
not well satisfied. We remark that the fidelity can be further improved by
reducing the errors via optimizing the systematic parameters. As
demonstrated in Fig. 7, a high fidelity greater than $99\%$ can be achieved
for $\Delta _{1}/g_{1}\gtrsim 32$ and $g_{kl}=0.01g_{\max }$ when the
unwanted couplings of the cavities (and the pulse) with the $\left\vert
e\right\rangle \leftrightarrow\left\vert f\right\rangle $ transition of the
flux qubit and the dissipation of the system are negligible.

The maximum among the coupling constants $\left\{
g_{1},g_{2},g_{3},g_{1}^{\prime },g_{2}^{\prime },g_{3}^{\prime }\right\} $
is $2\pi \times 0.163$\ GHz, which is available in experiments because a
coupling strength $\sim 2\pi \times 0.636$ GHz was reported for a
superconducting flux qubit coupled to a 1D microwave cavity [86]. As an
example, consider $T=20$ $\mu $s. In this case, the decoherence times of the
flux qubit used in the numerical simulations are $20$\ $\mu $s $-$\ $40$\ $%
\mu $s. Note that decoherence time 70 $\mu $s to 1 ms for a superconducting
flux qubit has been demonstrated in experiments [87,88]. Hence, the
decoherence time of the flux qubit considered in the numerical simulation is
a rather conservative case. In addition, the crosstalk strength $g_{kl}$
used in the numerical simulation can be obtained by prior design of the
coupling capacitances $C_{1},C_{2},C_{3}$ between the cavities and the
coupler qubit [89].

With the detunings $\Delta _{i}$ and the coupling constants $g_{i}$ listed
in Table 1 ($i=1,2,3$), one finds the effective coupling constant $\lambda
=g_{i}^{2}/\left\vert \Delta _{i}\right\vert \sim 2\pi \times 7.56$ MHz,
which results in $\Omega _{p}/2\pi \sim 1.89$ MHz for $s=1$ and $k=2$ used
in the numerical simulation. According to Eq. (18), the operation time is
given by $t=s\pi /\Omega _{p}.$ For $s=1$ and $\Omega _{p}/2\pi \sim 1.89$
MHz, a simple calculation gives the operation time $t\sim 0.26\ \mu $s, much
shorter than the qubit decoherence time applied in the numerical simulation
and the cavity decay time $10$\ $\mu $s $-$\ $100$\ $\mu $s considered in
Fig. 6. For $\kappa ^{-1}$ $=35$ $\mu $s and the cavity frequencies given
above, the quality factors for the three cavities (1,2,3) are respectively $%
Q_{1}\sim 9.89\times 10^{5},$ $Q_{2}\sim 8.18\times 10^{5},$\ and $Q_{3}\sim
6.59\times 10^{5},$ which can be achieved because a 1D microwave cavity or
resonator with a high quality factor $Q\gtrsim 10^{6}$\ was experimentally
demonstrated [90,91]. The analysis given above implies that implementation
of a three-qubit MCP gate using photonic qubits is feasible with the present
circuit QED technology.

\section{Conclusion}

We have proposed an efficient scheme to implement an $n$-qubit MCP gate,
i.e., a multiplex-controlled phase gate with $n-1$ photonic qubits
simultaneously controlling one photonic target qubit, based on circuit QED.
As shown above, this scheme is universal, because the two logic states of
each photonic qubit can be encoded via a vacuum state and an arbitrary
non-vacuum state $\left\vert \varphi \right\rangle $ (e.g., a Fock state, a
superposition of Fock states, a cat state, or a coherent state, etc.), which
is orthogonal or quasi-orthogonal to the vacuum state. In addition, this
scheme is simple because it requires only one step of operation.

This scheme has additional distinguishing features: (i) Since only two
levels of the coupler are used, i.e., no auxiliary levels are utilized,
decoherence from the higher energy levels of the coupler is avoided; (ii)
The gate operation time is independent of the number of qubits, thus it does
not increase with the increasing number of qubits; and (iii) The gate
implementation is deterministic because no measurement is needed.

As an example, we have numerically analyzed the circuit-QED based
experimental feasibility of realizing a three-qubit MCP gate with each
photonic qubit encoded via a vacuum state and a cat state. This scheme can
be applied to implement the MCP gate using photonic qubits in a wide range
of physical system, which consists of multiple microwave or optical cavities
coupled to a two-level coupler such as a natural atom or an artificial atom
(e.g., a quantum dot, an NV center, or a supercoducting qubit, etc.).
Finally, it is noted that an $n$-qubit Toffoli gate of photonic qubits can
be realized, by applying the present scheme to implement an $n$-qubit MCP
gate of photonic qubits, plus a single-qubit Hadamard transformation on the
target qubit before and after the $n$-qubit MCP gate.

\section*{Acknowledgements}

This work was partly supported by the National Natural Science Foundation of
China (NSFC) (11074062, 11374083, 11774076, U21A20436), the Key-Area
Research and Development Program of GuangDong province (2018B030326001), the
Jiangxi Natural Science Foundation (20192ACBL20051).



\end{document}